\documentclass[12pt]{article}
\usepackage{epsf}
\usepackage{epsfig}
\usepackage{hyperref}

\makeatletter
\renewcommand\section{\@startsection {section}{1}{\z@}%
                                   {-3.5ex \@plus -1ex \@minus -.2ex}
                                   {2.3ex \@plus.2ex}%
                                   {\normalfont\large\bfseries}}

\renewcommand\subsection{\@startsection{subsection}{2}{\z@}%
                                     {-3.25ex\@plus -1ex \@minus -.2ex}%
                                     {1.5ex \@plus .2ex}%
                                     {\normalfont\bfseries}}

\makeatother

\long\def\symbolfootnote[#1]#2{\begingroup%
\def\thefootnote{\fnsymbol{footnote}}\footnote[#1]{#2}\endgroup}

\def\({\bigl(}
\def\){\bigr)}
\def\<{\langle\,}
\def\>{\,\rangle}

\begin{document}
\begin{titlepage}

\begin{center}

\vskip  .7in {\Large \bf Geometry of Grand Unification}
\vskip .6in{Cumrun
Vafa}
\vskip .6in{Jefferson Physical Laboratory, Harvard University,
Cambridge, MA 02138}

\end{center}

\vskip .8in

\begin{abstract}
\baselineskip=16pt  
Grand Unification of all forces has been a well motivated
paradigm for particle physics.  This subject has been recently
revisited in the context of string theory, leading to a
geometric reformulation of the idea of unification of forces.
The interplay between geometry and physics has led to
a natural resolution to a number of puzzles of
particle physics utilizing the
geometry of extra dimensions of string theory.   Here we review aspects
of these developments for a mathematical audience (based on talks given in honor
of Yau's 60th, Atiyah's 80th and Singer's 85th birthdays).
\end{abstract}
\end{titlepage}

\section{Introduction}
The wish to have a unified theory of all forces is a long time dream
of physicists.  This dream which is rooted in aesthetics beauty 
and simplicity of nature, found concrete experimental evidence pointing
to its validity in the mid 1970's.  The basic idea is
that symmetries can be broken, and what appears at long distances
as distinct forces, may at shorter distances, and at higher energies, be part
of a single force.  In other words the symmetry between the forces transforming
one to another becomes restored at higher energies.

At the longest distance scale we are familiar with two forces: gravitational
force, and electromagnetic force.  Gravitational force is geometrized by Riemannian
metric on spacetime with signature $(3,1)$  where the matter influences the curvature through
Einstein's equations and where the matter is influenced from the
Riemannian structure by following geodesic paths.
The electromagnetic force on the other hand is based on the gauge principle.
In particular the geometrical data corresponds to a line bundle over the
spacetime with the electromagnetic gauge field being identified with a $U(1)$ connection
for this bundle. Moreover matter fields correspond to section of some associated
vector bundle depending on their $U(1)$ representation (i.e. their {\it charge}).
The story gets more interesting when we probe the physics at
yet shorter distances.

\section{Standard Model and Gauge Symmetry Breaking}

At shorter distance scale we know of two other forces:  At a distance scale
of about $10^{-13}cm$ we find that there is a strong force among quarks
binding them into nucleons. This is again based on the gauge principle of $SU(3)$
with the gauge field being identified with the adjoint connection of $SU(3)$.  
Again the various matter fields are described by sections of various vector
bundles associated with specific representations of $SU(3)$.  The strong forces
do not have a trace at longer distance because they are so strong they confine
quarks into neutral combinations, and we cannot find a single quark by itself.
At yet shorter distances
of about $10^{-16}cm$ we encounter the weak forces, responsible
for radioactive phenomena (which in particular can convert neutrons into protons).
This is again based on gauge principle, but this time it is the $SU(2)$ gauge field.
More precisely, the electromagnetic and weak forces correspond to the $SU(2)\times U(1)$
gauge field where the electromagnetic $U(1)$ sits in a diagaonal combination of
$U(1)_{em} \subset SU(2) \times U(1)$.  The main point is that the $SU(2)\times U(1)$
symmetry is `broken' to a diagonal $U(1)_{em}$ at larger distances. 

The notion of symmetry breaking has been a cornerstone of various
developments of the past few decades in theoretical physics.
In the context of gauge symmetry what this means is the following:
Suppose we have a matter field $H$  (called the `Higgs field') transforming
in a non-trivial representation of the gauge group $G$.  Suppose in
the vacuum the expection value of $H$ is not zero\footnote{More precisely
the gauge invariant $|H|^2$ has a vacuum expectation value.}.
Then we say the gauge symmetry is broken to a subgroup $K\subset G$
which preserves $H$.  This in particular means that the Green's functions
for the gauge fields in $G/K$ directions are not power law, but rather have an exponential
fall off set by the inverse scale of $H$.  This is known as the Higgs
mechanism, and is responsible for the breaking of the $SU(2)\times
U(1)$ to $U(1)_{em}$.

Thus at distance scales shorter than $10^{-17}cm$ we effectively
have a bigger gauge group, namely
$$G=SU(3)\times SU(2)\times U(1).$$
This is the gauge symmetry of the {\it standard model} of particle
physics.  The matter fields do not form a simple representation
under this group.  In fact they transform according to the following
highly reducible representation:
$$(3,2)_{-1}\oplus ({\overline 3},1)_{4}\oplus ({\overline 3},1)_{-2} \oplus (1,2)_{3}\oplus (1,1)_{-6}$$
The notation is that the $U(1)$ representation is denoted
by the subscript, and the two entries in the parenthesis correspond
respectively to dimension of $SU(3)$ and $SU(2)$ representations 
(the bar denotes complex conjugate representation).
These representations look somewhat complicated.  To make
the matters worse, they come in three copies.  In other words
we have to take the tensor product of this representation with
a 3 dimensional trivial representation $\otimes V$,
where $dim_{\bf C}V=3$.  These 3 copies of the
matter fields are called the {\it flavors}.  
In the next section we discuss some basic facts
known about flavors.

\section{Flavors and Hierarchy}
It is known experimentally that even though flavors come in three
copies, there is a way to distinguish the three flavors:  It turns
out that the masses of these flavors are very different and hierarchic.
The masses arise from {\it Yukawa} couplings, which corresponds
to cubic terms in the action given by
$$\lambda_{ij}\cdot [\psi_{M_i} \psi_{M_j} H]$$
where $H$ is the Higgs field and comes in pairs:  in the representation $(1,2)_{-3}$
for the up quarks and $(1,2)_{3}$ for the down quark.
 $i,j$ run over the three flavors and $\lambda$ is a $3\times 3$
matrix with suitable choice of matter fields $\psi_{M_i}$. For example
for the up-type quarks we choose the two matter fields to be 
$$q_L=(3,2)_{-1}, u_R=({\overline 3},1)_{4}.$$
with the Yukawa coupling
$$\lambda^u \cdot [q_Lu_R H_u].$$
 The three masses are obtained by considering eigenvalues
of the $3\times 3$ matrix $\lambda^{u\dagger} \lambda^u$.
We use  a unitary matrix to diagonalize this matrix
$$U_u \lambda^{u\dagger} \lambda^u U_u^{-1}=D_u^2$$
Similarly for the down type quarks we use the matter representations
$$q_L=(3,2)_{-1}, d_R=({\overline 3},1)_{-2}$$
and the three masses are given by considering eigenvalues of
 $\lambda^{d\dagger} \lambda^d$ with a unitary matrix
diagonalizing it denoted by $U_d$.

There are two facts about flavor physics which needs
an explanation:  The first one is that the mass eigenvalues are hierarchic.
For example for the up-type quarks the three masses are 
given by $(1.7,0.013,0.00003)\times 100 GeV$.
Typically this is `explained' by assuming that the corresponding
matrices have a Froggatt-Nielsen hierarchic entries given by
$$M_{ij}\sim a_{ij}\epsilon_1^{i-1} \epsilon_2^{j-1}$$
where $a_{ij}$ of are order 1 and $\epsilon_i$ are small parameters.
In addition to the mass hierarchy the other intriguing fact is that $U_u$ and $U_d$ are
very close. The almost basis independent object is
the unitary matrix given by
$$U_{CKM}=U_uU_d^{-1}$$
known as the CKM matrix.  It turns out that
$U_{CKM}$ is very close to the identity matrix. 
It is not difficult to see that up to a choice of 6 phases
for the basis of the
$u$ and $d$ quarks minus an overall phase rotation,
which does not affect the CKM matrix, the unitary
matrix is parameterized by $9-5=4$ parameters.
This can be chosen to be 3 real parameters and one phase.
The phase makes the unitary matrix not real which leads to violation of
complex conjugation symmetry (the CP symmetry
in physics terminology) and this has been experimentally observed
to be the case.   The phase is of order 1, however
the entries of the CKM matrix are very hierarchic.
  In 
fact if we consider the absolute value of the
entries of the CKM matrix it is given by
$$|U_{CKM}|\sim \left(\matrix {0.97&0.23&0.004\cr 0.23&0.97&0.04\cr 0.008&0.04&0.99\cr }\right )$$
Clearly these facts are in need of some explanation, and
standard model physics does not have a satisfactory explanation
of these structures.

\section{Unification of Gauge Groups}

Since the standard model is not a simple
group and the representations of matter fields are so complicated, one is naturally
led to ask:  Is there a bigger gauge group which is simple and includes
all the rest of the groups, such that the matter fields are in simpler
representations?  In such a case one can speculate that
the gauge symmetry gets enhanced at yet shorter distance
scale (higher energy scales).  The answer turns out to be
yes.  There are some choices.  The most minimal one
is the Georgi-Glashow $SU(5)$ model where the embedding
$$SU(3)\times SU(2)\times U(1)=S(U(3)\times U(2))\subset SU(5)$$ 
is the canonical one.  Furthermore the matter representation simplify:
The complicated matter representations we mentioned unify just
to two representations: The 10 dimensional anti-symmetric rank 2 representions and the 5 diemensional conjugate of the fundamental
representation!  This is remarkably simple.  The Higgs fields are
in $5\oplus {\overline 5}$ and the Yukawa couplings giving
mass to the up quarks come from
$$10_M\cdot 10_M \cdot 5_H$$
and for the down quarks from
$$10_M \cdot {\overline 5}_M \cdot {\overline 5}_H$$

  There are other choices
of unifications.  For example if the gauge group unifies to $SO(10)$
(by a further canonical embedding of $SU(5)\subset SO(10)$), then
these representations also unify to the 16 dimensional spinor
representation!  Indeed
$$16\rightarrow 10+{\overline 5}+1.$$
This gives an extra matter which transforms as trivial
representation of $SU(5)$ (and can be identified with right-handed
nuetrinos if they exist).

For the gauge factors to unify, their coupling constants $g_i$
should be the same.  This is because in the action we have only
one gauge invariant term
$${1\over g^2}{\rm tr} (F\wedge *F)$$
for the unified theory, and so if we read off the subgroup
which is identified with the standard model gauge group,
we have $g_i=g$.  The observed value
of the three gauge factors of the standard model $g_i$
are not equal.  However the coupling constants change with scale
due to quantum corrections.
It turn out that if we theoretically extrapolate the value of the coupling to
shorter distances (higher energy scales) where they have not
yet been experimentally measured, they come together (assuming
a supersymmetric completion of the standard model)
at a distance scale of about $10^{-30}cm$ (or an energy scale of about $10^{16}GeV$  (see
Fig. 1).  At that scale one finds
$$g_1=g_2=g_3=\sqrt{\alpha_{GUT}}\sim 0.2$$

 This is viewed as further evidence that the idea
of unification of gauge forces is correct.   This energy scale is still
much smaller than the Planck scale of $10^{19} GeV$, where
one expects quantum gravity effects to become dominant and
smooth spacetime loses its meaning.
If it had turned out that unification scale is at higher
energies than Planck scale, that would have meant the unification
never occurs, because energies above the Planck scale
are not physically meaningful.

There is another independent
fact pointing to this energy range, which has to do with neutrino
masses (whose review is beyond the scope of this paper). 
Putting all these evidences together, we see a convincing
case for unification of forces in nature at the GUT scale
of $10^{16}$ GeV.

\begin{figure}
\centering
\includegraphics[width=15cm]{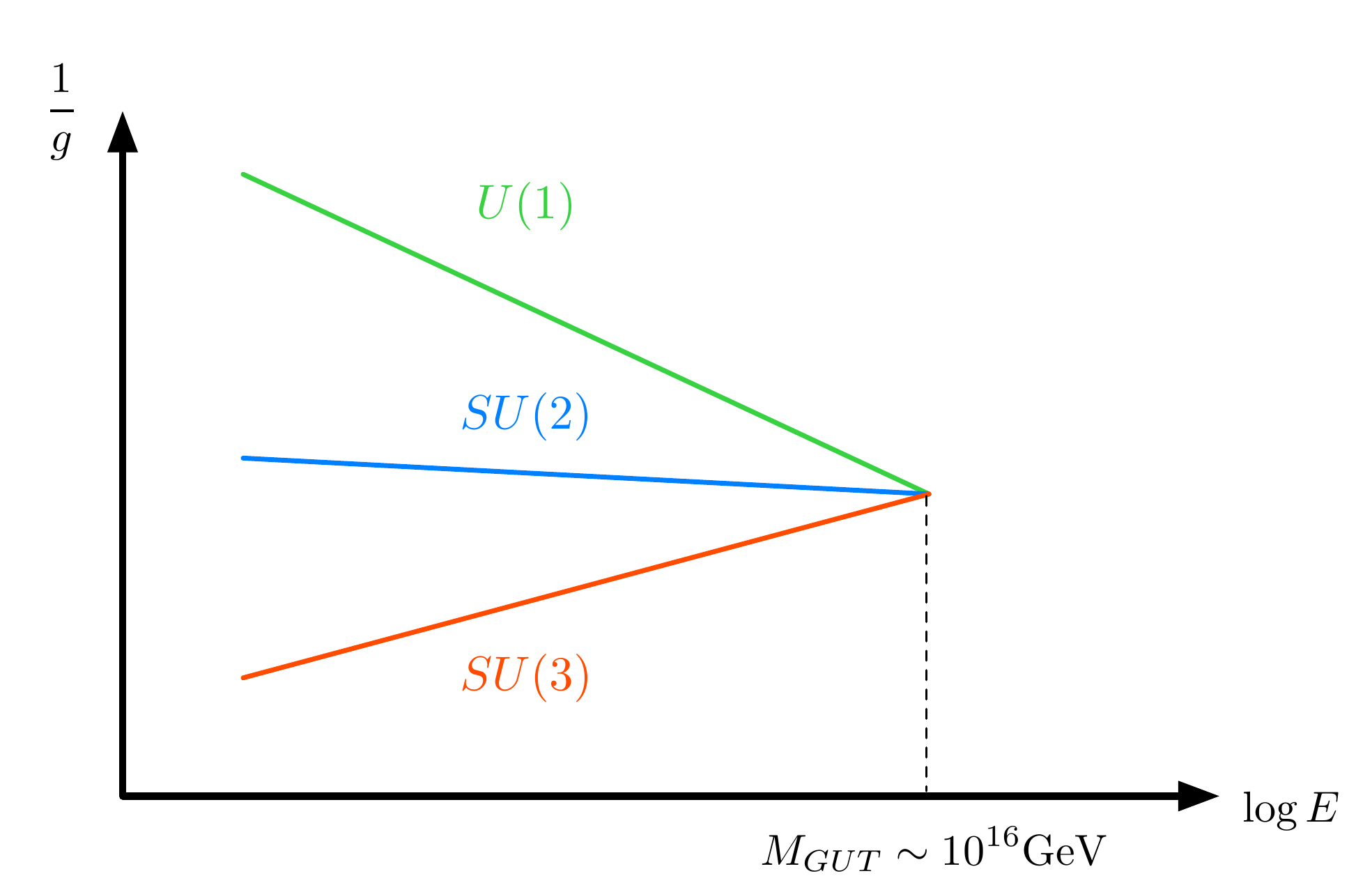} \setlength{\abovecaptionskip}{5pt}
\caption{{{ The couplings
of the three gauge groups unify at the scale of $10^{16}$ GeV. }}}
\setlength{\belowcaptionskip}{5pt}
\end{figure}

\section{String Theory, Forces, Matter, and Interactions}

String theory's main achievement in describing the real world
has to do with the fact that it provides a framework for a consistent
quantization of gravity.    However, it also naturally incorporates
gauge forces and matter, as well as interactions among them.
Geometry enters in a beautiful way in incorporating these
ideas:  It turns out that different objects can live in different
dimensions on subspaces of spacetime.  This is captured
by `branes' embedded in spacetime.
 Thus {\it geometrically
engineering of particle physics} by suitable choices of branes.
More precisely, the string vacuum corresponds to a geometry of the
form $R^4\times X$, where $R^4$ is the Minkowski space, and $X$ is
some compact manifold, and the brane can be embedded in
$R^4\times S$ where $S\subset X$.

String theory has a vast set of vacua (i.e. consistent choices
of $X$ and $S$).  This presents an
embarassment of riches!
In order to construct a stringy model for particular phenomenology
we need to know which vacuum out of this vast set corresponds
to our world.  In absence of a clear criteria to pick out this, we
cannot make progress in connecting string theory predictions with
observed particle physics data.

We will use one experimental hint to make progress:  The unification energy
scale  $10^{16}GeV << 10^{19}GeV$, the Planck scale.  The most
natural way to achieve this is to postulate that the unified gauge theory
(such as $SU(5)$) lives on a brane whose internal volume $S$ is much
smaller than the scales in $X$.  Mathematically this suggest that $S$ {\it
should be contractible inside} $X$.   This contractibility suggests
the existence of a (close to) vanishing cycle which mathematically is
very restrictive.  The idea would then be to try to describe the
local model of $X$ near $S$ and expect that the particle physics
data would only require the local data near $S$.  
To have the most amount of flexibility in particle physics
constructions, $S$ has to have the maximal dimension.  It turns
out that contractibility and this maximality in dimension of $S$ points
to a particular corner of string vacuum known as `F-theory', which
has been the subject of recent interest in connecting string theory
to particle phenomenology (see e.g.
\cite{BeasleyKW},\cite{BeasleyDC},\cite{DonagiCA})

\section{F-theory Vacua}

F-thoery vacua correspond to a strong coupling limit of type IIB strings
\cite{VafaXN}.
The geometry involved in constructing vacua in this setup is given by
a Calabi-Yau fourfold, which admits an elliptic fibration with a section.
Let $X$ denote this 3 complex dimensional section.  The physical
spacetime is identified with $R^4\times X$.  The data of the elliptic fibration over
$X$ encodes the `branes' in the F-theory setup.  In particular on complex 
codimension 1 loci  (4 real dimensional subspaces of $X$)
the elliptic fiber will have singularities.  The type of the elliptic fiber singularity
dictates what lives on the corresponding brane.  In particular for the A-D-E
type of singularity we obtain A-D-E gauge theory on the corresponding locus.
Thus to engineer for example an $SU(5)$ GUT theory, we would require
an $A_4$ elliptic singularity over a locus $S$ which is where the $SU(5)$ connection
lives.  In other words,  the theory has an extra
geometric ingredient:  the data of an $SU(5)$ bundle over
$R^4\times S$, which can lead to the standard model gauge group.

\subsection{Matter Fields}
Fields representing matter live on intersection of the loci
where elliptic fibration degenerates, i.e. on codimension 2 subspaces
corresponding to the intersection of two branes.  Let
$S_1,S_2$ denote two such branes, which correcspond
to A-D-E symmetries $G_1,G_2$.  In simple situations that
we will mainly focus, at the intersection
locus, the elliptic singularity type enhances corresponding to a group
$G_{12}$.  Let $\Sigma$ denote the curve where $S_1,S_2$ intersect:
$$S_1\cap S_2=\Sigma_{12}$$
Then the matter field that lives on $\Sigma_{12}$ is represented
by a section of a bundle associated to a representation $R_{12}$ of $G_1\times G_2$ obtained by adjoint decompistion of $G_{12}$ into
representations of $G_1,G_2$:
$$Adj(G_{12})\rightarrow Adj(G_1)\oplus Adj(G_2)\oplus R_{12}$$
More precisely such matter fields live on $R^4\times \Sigma_{12}$, where
$R^4$ represents the Minkowski space.  The explanation of this is that
locally near $\Sigma_{12}$ we can describe the geometry as a $G_{12}$ bundle data which is locally `Higgsed'.  This data
of Higgsing is captured by a $G_{12}$ adjoint valued field $\phi$ which captures
the unfolding of the elliptic singularity near $\Sigma_{12}$.
We can interpret the fact that we only have $G_1\times G_2$ gauge symmetry as due to the fact that these scalars $\phi$
have a (holomorphic) space dependent values which lead to breaking
of $G_{12}\rightarrow G_1\times G_2$ away from $\Sigma_{12}$.  We have a Hitchin like system with
equations given by
$${\overline \partial}_A \phi=0$$
$$F^{(0,2)}=0,$$
where $F$ is the curvature of the $G_{12}$ connection.
Note that these equations can be viewes as coming from an action
given by 
$$L=\int_S {\rm tr} (\phi \wedge  F^{0,2})=\int_S{\rm tr}(\phi\wedge ({\overline \partial}{
\overline A}+
{\overline A} \wedge
{\overline A})).$$
In other words we have a Hitchin like system (where
$\phi$ is a $G_{12}$ adjoint valued section of canonical bundle on $S$).  The
first order holomorphic deformations
of this local bundle data are the  matter fields and this deformation
is localized on the curve $\Sigma_{12}$ and given by representation $R_{12}$ given above.

For example, if we wish to have a particle in the representation $5$
of $SU(5)$ we need the $SU(5)$ brane to intersect a $U(1)$ brane
where on the intersection we get an enhancement to $SU(6)$.
The adjoint of $SU(6)$ decomposes to adjoint of $SU(5)\times U(1)$
and in addition the matter field in representation
$$R_{12}=5\oplus {\overline 5}$$
Similarly if we wish to obtain a matter field in the rank two antisymmetric
representation of $SU(5)$ we need an extra $U(1)$ brane and an intersection locus where the singularity type enhances to $SO(10)$.
The adjoint decomposition now leads to the matter representation
$$R_{12}=10\oplus {\overline {10}}$$
  To find which particles
they correspond to in 4-dimensions, we need to find the spectrum
of the Dirac operator on $\Sigma_{12}$:
$$D\psi_i =m_i \psi_i$$
The scale for the $m_i$ is set by the inverse size of $\Sigma_{12}$ which
is in turn set by the size of $S_i$.  Thus typically the $m_i$ have GUT
scale mass of order $10^{16} GeV$.  To find particles corresponding
to the one observed in nature (of the weak scale), we look for the ones which are massless in this limit.  In other words we look for zero modes
for the Dirac opertor.  The net number of
such modes (which is the number one expects to remain massless) is captured by the Atiyah-Singer index theorem:
$$ind(D)_R=\int_{\Sigma_{12}} F_R,$$
where $F_R$ denotes the curvature of the $G_1\times G_2$ bundle
over $\Sigma$ in the representions $R$ of matter fields
on $\Sigma$.  Thus we see a natural interpretation to the number
of flavors of the standard model:  The index is simply 3 for the
representations of matter fields.  In other
words
$$\int F_R=3$$
This makes the multiplicity of the matter fields much less exotic
and points to extra dimensions as the origin of this multiplicity.
 
\subsection{Yukawa Couplings}
Yukawa couplings arise in different ways, but the typical one involves
the triple intersection of branes.  In other words, it corresponds
to points in $X$ such that the singuarlity type is further enhanced.
Consider in particular three branes $S_i$, supporting gauge groups
$G_i$.  On the intersection of each pair of branes $S_i,S_j$ on the
curve $\Sigma_{ij}$ live the matter fields in represention
$R_{ij}$ of $G_i\times G_j$. On the point $p_{ijk}$ of triple intersection
the singulairty type enhances to $G_{ijk}$.  We have
$$G_i\subset G_{ij}\subset G_{ijk}$$
for all $i,j,k$.
Let $\phi_{ij}^\alpha$ corresponds to the zero modes
of the Dirac operator for matter fields on $\Sigma_{ij}$.  This leads to a Yukawa
coupling given by the product of the zero modes at the point of
intersection:
$$c_{\alpha \beta \gamma}= \phi_{ij}^\alpha (p_{ijk})\phi_{jk}^\beta (p_{ijk})\phi_{ki}^\gamma (p_{ijk})$$

In terms of the Hitchin-like system the Yukawa coupling is a measure
of the second order obstruction to deformation of the holomorphic Higgs bundle represented by the matter zero modes.  It turns out that this
leading computation of Yukawa coupling gets corrected
due to fluxes on $X$.  This will turn out to be important in the
applications which we will discuss below.

\section{Applications to Particle Physics}

We now discuss some simple applications of the ideas mentioned
above.  First, we show that the string geometry must include
a singularity of E-type at some point in the internal geometry.
Next we show that the flavor hierarchy can naturally be incorporated
in this setup.  Finally we conclude by noting how the standard
model gauge group arises in this set up.

\subsection{E-type Singularity}
Let us consider the minimal GUT theories, namely the unification in
$SU(5)$.  We need matter fields corresponding to representations
$10,{\overline 5}$, which means that on a curve we have an 
enhancement
$$SU(5)\rightarrow SO(10)\qquad for \quad 10$$
$$SU(5)\rightarrow SU(6)\qquad for \quad {\overline 5}$$
Similarly Higgs field is in the $5$ and ${\overline 5}$  over which we get an $SU(6)$
enhancement.
In addition we need to have the Yukawa interaction between
the matter fields and the Higgs.   In particular at one point we need
to have an enhanced symmetry group to get
$$10_M\cdot 10_M\cdot 5_H$$
for the top quarks and
$$10_M \cdot {\overline 5}_M \cdot {\overline 5}_H$$
for the down quarks and leptons.
The top quark mass interaction implies that at the intersection
point we have a further enhancement:
$$SU(5)\rightarrow (SU(6),SO(10))\rightarrow E_6\qquad top \quad quark$$
$$SU(5)\rightarrow (SU(6),SO(10))\rightarrow SO(12)\qquad down \quad quark$$
We will later argue that these two points of enhanced singularity should
be very close to each other, in order to explain the hierarchy
in the CKM matrix.  Bringing these two points together, i.e.
combining the $E_6$ and $SO(12)$ symmetries
leads to the yet higher symmetry $E_7$ at the intersection point.
Moreover the requirement that supersymmetry breaking is
visible only at a very low scale, requires an extra rank at
the intersection point, leading to $E_8$ symmetry point.  This
is quite remarkable!  Simply trying to accomodate what
we know for observed particles and their flavor structure,
and guided by the principle of unification of forces and embedding
into string theory we are automatically
led to $E_8$ symmetry \cite{HeckmanMN}!  This is indeed a rich interplay between
particle physics and geometry.

\subsection{Flavor Hierarchy}

As mentioned before the matter comes in three copies.  This
multiplicity can simply be accomodated as zero modes of the Dirac 
opertors.  However there is more to the flavor structure:  Their
masses are very hierarchic.  This means that the corresponding
Yukawa matrix is hierarchic.  For example for the down quarks
we have
$$c_{\alpha\beta}10^\alpha_M{\overline 5}^\beta_M {\overline 5}_H$$
where $\alpha,\beta=1,2,3$,
and we need a hierarchic matrix $C=c_{\alpha \beta}$, namely the
eigenvalues of $CC^\dagger$ should be very hierarchic.  In
the context of F-theory, we can compute $C$, as already
noted.  In the limit we ignore fluxes this is simply given by the
multiplication of the zero modes living on each curve, at the joint
intersection point $p$:
$$c_{\alpha \beta}= \phi_{10}^\alpha (p)\phi_{{\overline 5}}^\beta (p)\phi_{\overline 5}(p)$$
Note that this $3\times 3$ matrix has at most rank one (because
it is given by outer product of the two vectors, $\phi_{10}^\alpha (p)$
and $\phi_{{\overline 5}}^\beta (p)$.  It is useful to choose
a basis of zero modes adapted to order of vanishing at $p$.
Namely on the curve supporting $\phi_{10}^\alpha (p)$ near
$p$ choose a coordinate chart $z_1$, with $z_1(p)=0$.  Similarly 
on the curve supporting  $\phi_{{\overline 5}}^\beta (p)$ choose coordinate
chart $z_2$ near $p$ with $z_2(p)=0$.  In this way a basis for the zero modes can be chosen to go like $(1,z_1,z_1^2)$ and $(1,z_2,z_2^2)$ 
near the intersection point, and thus the
matrix $C$ has only one non-zero entry.  In this case we have
the extreme flavor hierarchy, where we have two massless flavors
and one massive.  However, when we turn on flux this changes
depending on the choice of flux \cite{HeckmanQA}.  Mathematically turning on 
fluxes correspond to making the Hitchin-like system live on a non-commutative space \cite{Cecotti:2009zf}.
This makes the flavaor mass matrix hierarchic.  In this case the local
$U(1)\times U(1)$ phase rotation symmetry of the $z_1,z_2$ 
imposes order of symmetry violation on the matrix elements of $C$.
The non-vanishing fluxes leading to non-commutativity,
correspond to breaking this rotation phase
symmetry and thus impose hierarchic structure on the $c_{ij}\sim
\epsilon_1^{i-1}\epsilon_2^{j-1}$.
This explains the natural geometrization of flavor hierarchy in string theory.

Another aspect of flavor hierarchy is the fact that the CKM mixing
matrix between u-type quarks and d-type quarks is hierarchic.
This question can also be geometrized beautifully in the context
of F-theory GUT models:  On the curve $\Sigma_{10}$ where
the three $10$ matter field zero modes live, there are two special points:  From one of them $q$, 
we get the u-type quark masses (the 
$10\cdot 10\cdot 5$ interaction) and the other point $p$, the d-type
quark masses ($10\cdot {\overline 5}\cdot {\overline 5}$).
The corresponding mass matrices are diagnoalized in a different
basis of zero mode wave function. The CKM mixing matrix
is the unitary matrix which takes one basis to the other.
Apriori there is no reason this matrix is close to identity,
as is experimentally observed.  However, there is apriori
no reason that this should be so.  However, if we assume $p=q$
then the basis vectors which diagonalize both of the mass matrices are close to the basis
given by the order of vanishing of the wave function at that point.
Thus the unitary matrix which takes one to the other is close
to the identity.  Using estimates of this rotation ones gets an
estimation of this unitary matrix \cite{HeckmanQA} :
$$|U^{F-theory}_{CKM}|\sim \left(\matrix {1&\epsilon &\epsilon^3\cr \epsilon&1&\epsilon^2\cr \epsilon^3&\epsilon^2&1\cr }\right ) $$
where $\epsilon \sim \sqrt{\alpha_{GUT}}\sim 0.2$,
 in good rough agreement with
the experimentally observed matrix.

Note further that the requirement of $p=q$ enhances
the symmetry by combining the $E_6$ and the $SO(12)$ enhancement
points to $E_7$ (and ultimately to $E_8$ as noted before).

\subsection{Breaking to the Standard Model}
We have discussed how the $SU(5)$ symmetry arises
geometrically by having an $SU(5)$ elliptic singularity over
the brane $S$.  In order to obtain the standard model
gauge group $S(U(3)\times U(2))$ we need to break this gauge
group.  It turns out that can be simply done by having a curvature
in a $U(1)$ sub-bundle of $SU(5)$.  The $U(1)$ direction
embeds in the Cartan of $SU(5)$ in the direction $(2,2,2,-3,-3)$.
Moreover the curvature one needs to choose to solve string
equations leads to anti-self dual configuration (i.e.
a $U(1)$ instanton) on $S$.  The sub-bundle which preserves
this structure is the $S(U(3)\times U(2))$ which thus emerges
as the gauge symmetry in 4 dimensions.  This is the desired symmetry
of the standard model.  

There is one interesting geometric subtlety in this breaking.  Namely
to make sure the $U(1)$ of the standard model is not broken
by this flux, we need the flux that we turn on over $S$ to be dual
to a 2-cycle, which is contractible inside $X$.

\section{Further Issues}

In this paper we have reviewed some of the recent developments
in reformulating particle physics in geometric terms in the context
of string theory and using some features of geometry to explain
some of the puzzles of particle physics.  In trying to make this
link stronger a number of mathematical issues need to be better
understood:  The geometry of vanishing 4-cycle supporting
the GUT group in the base
of elliptic Calabi-Yau fourfolds plays a key role.  One needs
to study aspects of this and find what restrictions this puts
on the geometry (see in particular 
\cite{DonagiRA},\cite{clay}).  

In addition to this it has been found that monodromy of the branes
plays a key role in understanding of phenomenology (see
in particular \cite{HayashiGE}.  What
this means is that the loci of elliptic singularity, which can be
formulated in terms of spectral covers undergoes monodromy.
It would be important to understand this geometry more precisely
and also more deeply follow its interplay with contractibility
of the 4-cycle supporting the GUT brane.

\vglue 2cm

I would like to thank Jonathan Heckman and my other collaborators on the F-theory GUT model
building for very exciting collaborations.
This  research was supported in part by NSF grant PHY-0244821.

\end{document}